\documentclass[%
superscriptaddress,
nobibnotes,
 amsmath,amssymb,
 aps,
 prl,
 twocolumn,
]{revtex4-2}

\usepackage[english]{babel}
\usepackage{graphicx}
\usepackage{dcolumn}
\usepackage{bm}
\setcitestyle{super}

\usepackage{verbatim}
\usepackage{siunitx}
\usepackage{hyperref}
\usepackage{nicefrac}

\begin{document}

\newcommand{\UniLu}{University of Luxembourg, Department of physics and materials science, 41 rue du Brill, 4422 Belvaux, Luxembourg}
\newcommand{\LIST}{Materials Research and Technology Department, Luxembourg Institute of Science and Technology (LIST), 41 rue du Brill, 4422 Belvaux, Luxembourg}
\newcommand{\ULiege}{Quantum Materials Center (Q-MAT), Complex and Entangled Systems from Atoms to Materials (CESAM), Universit\'e de Li\`ege, Quartier Agora, All\'ee du Six Aout, B-4000 Li\`ege, Belgique}
\newcommand{\Neel}{Institut N\'eel CNRS/UGA UPR2940, 25 Rue des Martyrs, 38042 Grenoble, France}
\newcommand{\Soleil}{Synchrotron SOLEIL, L’Orme des Merisiers, Départementale 128, 91190 Saint-Aubin, France}
\newcommand{\CNR}{CNR-SPIN, c/o Universit\`a ``G. d'Annunzio", Chieti, Italy}
\newcommand{\ILM}{Institut Lumi\`ere Mati\`ere (ILM)-UMR 5306, Universit\'e Claude Bernard Lyon 1, Campus LyonTech-La Doua, 10 rue Ada Byron, F-69622 Villeurbanne, France}
\newcommand{\ICMCB}{Institut de Chimie de la Mati\`ere Condens\'ee de Bordeaux (ICMCB)-UMR 5026, CNRS, Universit\'e de Bordeaux, 87 Avenue du Docteur Schweitzer, F-33608 Pessac, France}
\newcommand{\ESRF}{ESRF -- European Synchrotron Radiation Facility, 38000 Cedex Grenoble, France}
\newcommand{\ALBA}{CELLS -- ALBA Synchrotron Light Source, Cerdanyola del Valles, Barcelona E-08290, Spain}
\newcommand{\ICPMS}{Universit\'e de Strasbourg, CNRS, Institut de Physique et Chimie des Mat\'eriaux de Strasbourg, UMR 7504, 67000 Strasbourg, France}

\newcommand{\KNO}{KNbO$_3$}


\title{An underdog story: Re-emergence of a polar instability at high pressure in KNbO$_3$}

\author{Mohamad Baker Shoker}
\thanks{These authors contributed equally to this work.}
\affiliation{\UniLu}
\author{Sitaram Ramakrishnan}
\thanks{These authors contributed equally to this work.}
\affiliation{\Neel}
\author{Boris Croes}
\author{Olivier Cregut}
\author{Nicolas Beyer}
\author{Kokou D. Dorkenoo}
\affiliation{\ICPMS}
\author{Pierre Rodi\`ere}
\affiliation{\Neel}
\author{Bj\"orn Wehinger}
\author{Gaston Garbarino}
\author{Mohamed Mezouar}
\affiliation{\ESRF}
\author{Marine Verseils}
\author{Pierre Fertey}
\affiliation{\Soleil}
\author{Salia Cherifi-Hertel}
\affiliation{\ICPMS}
\author{Pierre Bouvier}
\affiliation{\Neel}
\author{Mael Guennou}
\email[]{mael.guennou@uni.lu}
\affiliation{\UniLu}

\date{\today}

\begin{abstract}
Ferroelectricity in perovskites is known to be suppressed by a moderate hydrostatic pressure. The notion that a polar instability should reappear in a higher pressure regime is well accepted theoretically but experiments have failed so far to provide a conclusive evidence for it. Here, we investigate a classical but comparatively underlooked ferroelectric perovskite KNbO$_3$. We use single crystal X-ray diffraction, infrared and Raman spectroscopy and second-harmonic generation to explore the phase transition sequence at high pressures up to \SI{63}{\GPa}. We show that the ferroelectric instability manifests itself in the emergence of an incommensurate modulation of the perovskite structure that combines cation displacements and tilts of the oxygen octahedra. Soft modes associated to the tilts and the modulation are clearly observed along with persistent order-disorder signatures. This demonstrates the presence of the high-pressure polar instability in a lead-free perovskite in spite of the centrosymmetric character of all observed high-pressure phases.
\end{abstract}

\maketitle

Structural instabilities and distortions in perovskite oxides $AB$O$_3$ are fundamental to the understanding of their physical and functional properties. They are commonly decomposed in different types based on their symmetry: tilts of the $B$O$_6$ octahedra, (anti)polar displacements of the $A$ and $B$ cations and distortions of octahedra. As part of a general effort to understand these instabilities and their interplay in detail, perovskites have been studied under hydrostatic pressure with the aim of formulating general rules governing the behavior of their instabilities, isolated and in combination. For a long time, notably following pioneering work by Samara {\it et.~al.}~\cite{Samara1975}, the common understanding was that those rules were simple: ferroelectric soft modes at the Brillouin zone center should harden under hydrostatic pressure, while soft tilt modes at the zone boundary should soften. Over time, it was realized that the reality is more complex. Tilt instabilities can be in fact either enhanced or suppressed by pressure, as it was confirmed experimentally~\cite{Bouvier2002} and subsequently rationalized in several papers refining the general rules with a greater attention to the details of energetics and atomistic interactions~\cite{Angel2005,Zhao2004,Xiang2017}. As far as ferroelectric instabilities are concerned, it was checked that classical ferroelectric perovskites undergo a ferro- to paraelectric cubic phase transition~\cite{Pruzan2007}. In the early 2000s, these views were revisited and challenged by first-principle calculations~\cite{Kornev2005,Kornev2007} with the key prediction that the ferroelectric soft mode frequency indeed hardens at first, thereby causing a ferro- to paraelectric transition, but then re-soften in a much higher pressure regime, albeit with a drastic change in the phonon eigenvector~\cite{Bousquet2006}, which should lead to a re-entrance of ferroelectricity.

As the prototypical displacive ferroelectric perovskite with well-documented soft phonon modes at ambient pressure, lead titanate PbTiO$_3$ was the natural contender to check this prediction. Very early on, a combined XRD and Raman spectroscopy study on PbTiO$_3$ at high pressure was reported, claiming evidence for this new high-pressure ferroelectric phase above \SI{50}{\GPa}~\cite{Janolin2008}. However, the evidence came from powder diffraction alone, was particularly fragile and could never be confirmed. In a very recent study~\cite{Cohen2024}, second-harmonic generation (SHG) was performed and gave no evidence for a breaking of inversion symmetry up to \SI{90}{\GPa}, leading the authors to conclude on the absence of a high-pressure ferroelectric ground state. Other attempts in other candidate systems in similar pressure ranges (SrTiO$_3$, CaTiO$_3$ etc.)~\cite{Guennou2010,Guennou2010a} have been equally unsuccessful in providing a conclusive proof of this high-pressure ferroelectric soft mode. While this does not invalidate the well-accepted theoretical views, it suggests that re-entrance of ferroelectricity might only happen at much higher pressures where they would have to compete with transition to post-perovskite phases~\cite{Cohen2024} -- and where experiments are even more challenging. 

In this letter, we investigate the high-pressure behavior of potassium niobate KNbO$_3$. KNbO$_3$ is a classical, lead-free ferroelectric perovskite. It is isostructural to BaTiO$_3$ and is known as a dominantly order-disorder type ferroelectric, where some phonon softening is present but incomplete~\cite{Fontana1984}. Its $P$-$T$ phase diagram up to the paraelectric cubic phase was mapped in details~\cite{Pruzan2007}. The persistence of a Raman signal~\cite{Pruzan2007} and diffuse x-ray scattering up to at least \SI{26}{\GPa}\cite{Ravy2007} indicate that Nb local disorder persists even in the cubic phase, which does not make it a priori a favorable ground when looking for ferroelectric soft modes. Yet we will show that the ferroelectric instability indeed reappears under pressure, but has to coexist with the tilt instability that is also favored by pressure. 

\begin{figure}[th]
    \centering
   \includegraphics[width=0.5\textwidth]{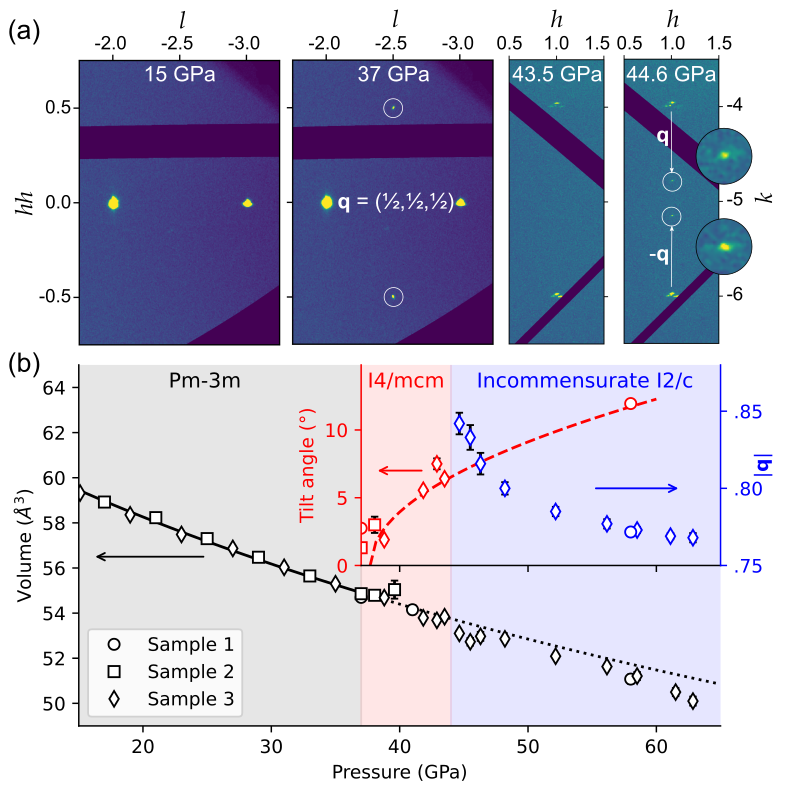}
    \caption{(a) Reciprocal space reconstructions at selected pressures. Circles indicate the positions of the superlattice reflections, commensurate and incommensurate. The reconstructions at 15 and \SI{37}{\GPa} are in the $hhl$ plane with respect to the cubic lattice and the maps at 43.5 and \SI{44.6}{\GPa} in the $hk\bar 1$ with respect to the tetragonal lattice. (b) Volume, tilt angle and modulation $|\mathbf q|$ as a function of pressure. The solid line is a fit to a 2\textsuperscript{nd}-order Birch-Murnaghan equation of state on the cubic volume and the dotted line is its extrapolation. The dashed line for the tilt angle is a guide to the eye.}
    \label{fig:tetra_incomm_unwarp}
\end{figure}


KNbO$_3$ single crystals grown by the top-seeded solution growth were purchased from SurfaceNet GmbH (Germany) for a series of high-pressure experiments. Experimental details are given in the Supplemental Material\cite{SupplMat} (See also Refs.\cite{Shen2020,Takemura2001,Dewaele2007,crysalis} therein). All results in the low-pressure range across the ferroelectric to paraelectric transition were found essentially consistent with the literature~\cite{Pruzan2007} and will not be shown here. High-pressure single-crystal X-ray diffraction at room temperature were carried out at the SOLEIL synchtrotron source (CRISTAL beamline) and at the ESRF (beamlines ID27~\cite{Mezouar2024-ID27} and ID15B~\cite{Garbarino2024-ID15B}). At \SI{37}{\GPa}, we observe the appearance of superlattice reflections at commensurate positions \textbf{q} = $(\frac{1}{2}, \frac{1}{2}, \frac{1}{2})$ (labelled as the $R$ point of the cubic Brillouin zone) as shown in Figure \ref{fig:tetra_incomm_unwarp}. All diffraction maxima were indexed on an $I$-centered lattice and the structure was successfully described by the space group $I$4$/mcm$ using the CrysalisPro software package \cite{crysalis} and Jana 2020 \cite{petri2023a}. 
It corresponds to the transition to a structure with anti-phase tilts of oxygen octahedra ($a^0a^0c^-$ in Glazer notation~\cite{Glazer1975}). The presence of this phase has not be reported so far in KNbO$_3$, including in a recent powder diffraction study~\cite{Kobayashi2000}, but this occurrence is not surprising. It is pretty ubiquitous in many perovskites, ferroelectric or not (SrTiO$_3$~\cite{Guennou2010}, PbTiO$_3$~\cite{Janolin2008}, BaZrO$_3$~\cite{Toulouse2022}, EuTiO$_3$\cite{bes2018a}), as a result of a pressure-enhanced instability of this tilt mode at the $R$ point\cite{Xiang2017}. Complementary DFT calculations of the phonon dispersion (provided in the supplementary information) confirmed that the tilt phonon mode at the $R$ point becomes more and more unstable under pressure, and further preliminary results confirm the stabilization of this tetragonal phase under pressure~\cite{GhosezPC}.

\begin{figure*}[ht]
    \centering
    \includegraphics[width=\textwidth]{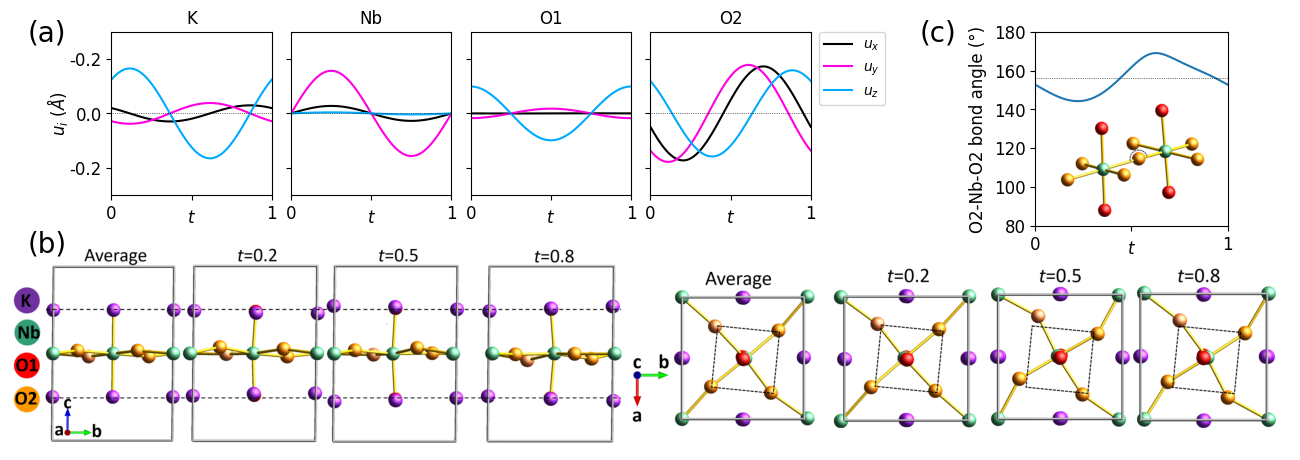}
    \caption{Atomistic depiction of the modulation at \SI{58}{\GPa} (a) $t$-plots showing all atomic displacements as a function of the phase of the modulation. (b) Projections of the crystal structure of KNbO$_3$ onto the \textbf{bc} and \textbf{ab} planes at selected $t$ values. Dashed lines serve as a reference to depict the displacement of the atoms from the average structure. (c) Nb-O-Nb bond angle as a function of $t$. The dashed horizontal line represents the bond angle for the average structure. }
    \label{fig:modulation}
\end{figure*}

At \SI{44.6}{\GPa}, we observed the emergence of incommensurate satellite reflections at \textbf{q} = 0.843(7)\textbf{b*} (Fig.~\ref{fig:tetra_incomm_unwarp}(a)). The incommensurate \textbf{q} was observed to shrink rapidly upon increasing pressure and level off at $\approx 0.77$. The superspace approach \cite{van2007incommensurate} was employed to elucidate the incommensurate modulation. Initial observation revealed satellites along \textbf{a*} and \textbf{b*} as shown in the Supplemental Material\cite{SupplMat}. However, (3+2)-dimensional ($d$) modulation was ruled out in favor of (3+1)-$d$ supported by absence of mixed-order satellites. Instead, the satellites along \textbf{a*} belong to another domain brought about by the loss of the 4-fold rotational symmetry around \textbf{c*}. Further evidence proving that the modulated phase is no longer tetragonal arises from the uniaxial direction of the wavevector that is incompatible with the tetragonal ($I$4/$mcm$) symmetry in (3+1)-$d$ but possible with its orthorhombic and monoclinic subgroups \cite{stokesht2011a}. From structural refinements using Jana 2020 \cite{petri2023a}, a better fit was obtained for a monoclinic superspace group $I$2/$c(0\sigma_1\sigma_2)0s$ ($a$-unique ($\sigma_2 = 0$) with $\alpha = \SI{89.40(10)}{\degree}$) as compared to its orthorhombic supergroup $Ibam(0\sigma 0)s00$. A detailed comparison for both models as well as other crystallographic information are provided in the Supplemental Material~\cite{SupplMat}. 

An atomistic depiction of the modulation at \SI{58}{\GPa} is given in Figure~\ref{fig:modulation}. It is first represented by so-called $t$-plots showing the amplitude of atomic displacements from the average structure as a function of the phase of the modulation $t$. Fig.~\ref{fig:modulation}(b) shows a pictorial visualization of the modulation by comparing the average structure to the modulated structure at a few selected $t$ values. In addition, a comprehensive visualization of the modulated structure is provided in the form of a movie generated with Jana 2020 \cite{petri2023a, SupplMat}. The modulation appears as quite complex and involves a combination of several distortions. Both K and Nb cations are displaced from their high-symmetry position, thereby breaking inversion symmetry locally. However, Nb is displaced longitudinally along \textbf{b}, i.e. along the modulation vector, while K (as well as O1) undergo a transverse modulation along \textbf{c}. The basal O2 ions appear to be the most heavily modulated in all directions. This can be seen as an \emph{in-phase} tilting motion along the \textbf{c} axis that superimposes to the \emph{anti-phase} tilting defining the average $I4/mcm$ structure. This results in a significant modulation of the Nb-O2-Nb bond angle between 169.0(3)$^\circ$ and 144.3(4)$^\circ$ (Fig.~\ref{fig:modulation}(c)). Finally, octahedra also undergo stretching along the \textbf{c} direction that accompany the K displacements. 

\begin{figure}[ht]
    \centering
    \includegraphics[width=0.5\textwidth]{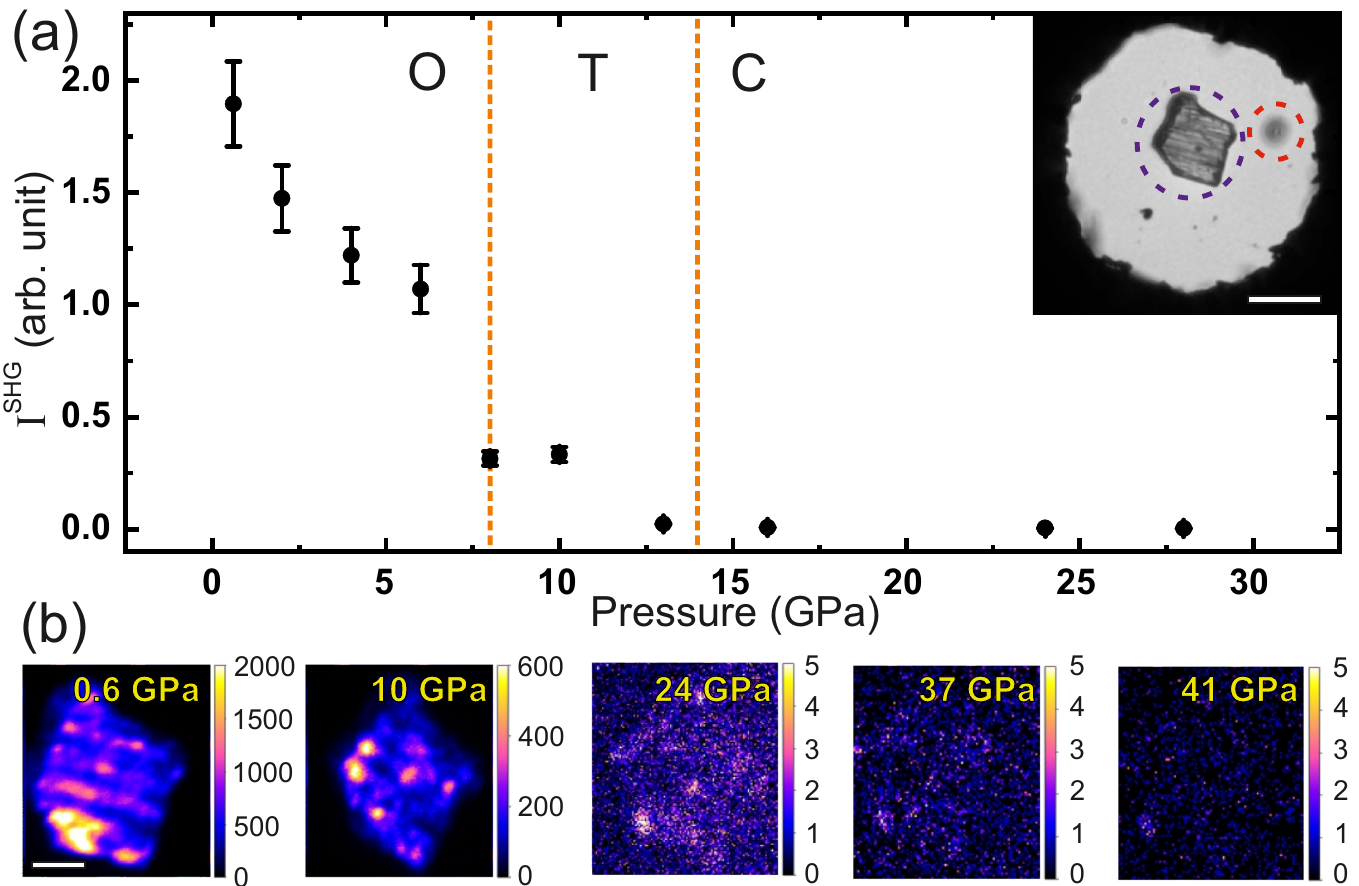}
    \caption{Variation of the second-harmonic generation (SHG) response with pressure. (a) Overall isotropic intensity variation. Dashed lines separate pressure ranges corresponding to the different phases. The inset displays an optical microscopy image showing the sample (blue dashed circle) and the ruby ball (red dashed circle). (b) SHG microscopy images of the sample at selected pressures. The scale bar is \SI{10}{\um}.}
    \label{fig:shg}
\end{figure}

It is sometimes argued that conventional XRD cannot unambiguously determine the presence (or breaking) of macroscopic inversion symmetry and that a dedicated experimental method such as second-harmonic generation (SHG) should be used instead~\cite{Cohen2024}. Indeed, in spite of the use of parameters like Flack, Parsons etc. in modern crystallographic softwares that does allow to distinguish between centrosymmetric and non-centrosymmetric periodic crystals, aperiodic structures can remain problematic. This was illustrated recently with the incommensurate modulation in EuAl$_4$ where the breaking of inversion symmetry was overlooked in a first XRD study~\cite{ramakrishnan2022a} before being proposed by SHG measurements \cite{yang2024a} and finally corroborated by further diffraction experiments~\cite{kotla2025b}. With this in mind, we performed SHG imaging on a KNbO$_3$ crystal up to \SI{58}{\GPa}. Details are described in the Supplemental Material\cite{SupplMat} and Refs.~\cite{Cherifi-Hertel2021, Croes2022} therein. The details of the pressure-dependent SHG measurements are described in the supplementary information. As shown in Fig.~\ref{fig:shg}, the SHG intensity averaged over the sample area systematically decreases with increasing hydrostatic pressure. Three distinct regimes can be identified, each corresponding to different phases: $Amm$2 (orthorhombic), $P$4$mm$ (tetragonal), and ultimately centrosymmetric $Pm\bar 3m$ (cubic). The corresponding SHG images show a clear domain structure with modulated intensity. At \SI{14}{\GPa}, the intensity drops to essentially zero and didn't show any sign of increase up to the maximum pressure of \SI{58}{\GPa}. Note that the high-pressure cell used for this experiment was the same from which the XRD dataset from Fig~\ref{fig:modulation} was collected, so that there is no doubt that the crystal was in the incommensurate phase, then confirmed to be centrosymmetric. 


\begin{figure*}[ht]
    \centering
    \includegraphics[width=0.9\textwidth]{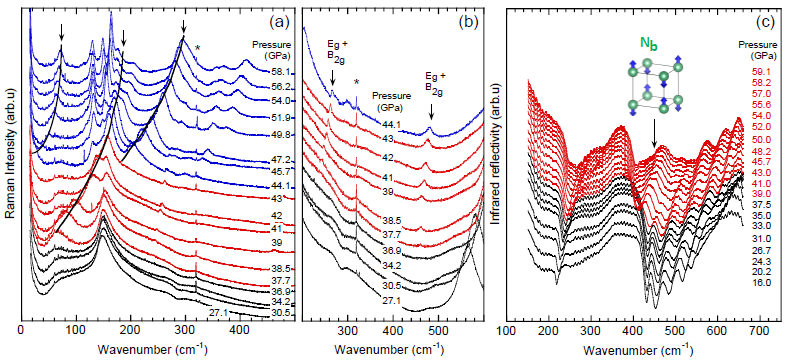}
    \caption{(a) Low-wavenumber Raman spectra showing the different soft modes as discussed in the text. (b) Raman spectra focused on the emergence of the $E_g+B_{2g}$ hard modes associated to the cubic-to-tetragonal transition. (c) Infrared reflectivity spectra. Black, red and blue colors refer to cubic, tetragonal and incommensurate phases respectively. The star * is an artifact.}
    \label{fig:Raman}
\end{figure*}

We now turn to the spectroscopic signatures of the transitions. Raman spectroscopy was performed in several experiments and up to a maximum pressure of \SI{58.1}{\GPa}. The transition to the cubic phase was observed at \SI{14.3}{\GPa}. Even though no first-order spectrum is allowed by symmetry in a cubic perovskite, the spectrum observed for cubic KNbO$_3$ remains characterized by an intense and structured spectrum shown in the Supplemental Material\cite{SupplMat}. This is classically explained by the persistence of local disorder; the Nb$^{5+}$ highly charged $d^0$ cation is known to stay shifted out of the centers of their octahedra through a pseudo-Jahn–Teller distortion, which occurs when the empty $d$ orbitals form hybrid orbitals with the filled p orbitals of the ligands~\cite{Bersuker1966}. Interestingly, this disorder-induced Raman signal weakens but does not vanish with increasing pressure and is still very prominent at \SI{30}{\GPa} at the onset of the anti-ferrodistortive (AFD) transition, which points to a persistence of the local disorder. 

Figure~\ref{fig:Raman} (a) shows the Raman spectra across both high-pressure transitions. Between 37 and \SI{43}{\GPa}, we observe several changes that meet the well-documented expectations for the $Pm\bar 3m$ to $I4/mcm$ AFD transition~\cite{Guennou2010,Toulouse2022}. This includes the emergence of the soft tilt mode (Fig.~\ref{fig:Raman} (a)) and the activation of hard modes with $E_g+B_{2g}$ symmetry, the splitting of which is allowed by symmetry but negligible in practice~\cite{Toulouse2022}. The latter are weak in intensity against the disorder-induced signal, as highlighted in Fig.~\ref{fig:Raman} (b), but their positions match the phonon calculations and their assignment is unambiguous. A much more spectacular change in the Raman signature occurs at \SI{44}{\GPa} with the emergence of the incommensurate modulation. We observe the activation of many sharp hard modes (Fig.~\ref{fig:Raman}~(a)). In addition, two modes pointed by arrows in Fig.~\ref{fig:Raman}(a) show a pressure hardening akin to soft-mode behavior that we attribute to specific modes named amplitudon and phason that are expected to be activated by the incommensuration~\cite{Cummins1990}. The detailed mode assignment is beyond the scope of this work but we note that the general behavior is compatible with the idea of a change from a phonon density-of-states dominated scattering to a set of discrete lines resulting from the activation of phonons at specific $k$ points. Remarkably, we note that none of the observed soft modes exhibits complete softening, but rather appear at some finite frequency, which is in line with the behaviour of ferroelectric soft modes at ambient pressure and the generally accepted order-disorder character of KNbO$_3$.

Infrared reflectivity spectra were recorded in the same pressure range at the SOLEIL synchrotron source, on a dedicated setup on the AILES beamline~\cite{Voute2016}. The most prominent signature in the spectra shown in Fig.~\ref{fig:Raman}(c) is the activation of a mode located at about \SI{400}{\per\cm} at \SI{37}{\GPa}.  Based on calculated phonon frequencies (SI), we attribute this mode to the zone boundary $R_4^-$ mode involving antipolar motions of Nb. This is known and expected from the symmetry lowering from $Pm\bar 3m$ to $I$4$/mcm$ but is usually found to be very weak~\cite{Yazdi-Rizi2017}. We speculate that the unusual strength of this mode here may reflect the proximity of the incommensuration and the modulation of the Nb position. In contrast, the activation of the silent $T_{2u}$ (so-called "butterfly") mode of oxygens is not observed, while the weaker signature around \SI{300}{\per\cm} is assigned to the splitting of a polar $T_{1u}$ mode. The transition to the incommensurate phase itself was not observed in that experiment, which we attribute to comparatively poorer pressure condition. Incommensurate phases are known to be fragile against perturbations: suppression of the modulation due to structural defects like dislocations, stacking faults or vacancies has been reported in systems such as $R$Te$_3$ ($R$ = rare-earth)~\cite{siddique2024a} and CuV$_2$S$_4$~\cite{ramakrishnan2019a}. FTIR spectroscopy experiments typically require larger samples that are more susceptible to bridging in the pressure chamber, thereby causing a degradation in crystal quality and a suppression of the modulation. Measurements in this pressure range therefore remain a challenge for state-of-the-art FTIR under pressure. 


Incommensurate modulations in simple perovskites are rare but not unheard of. It was proposed that EuTiO$_3$ shows a modulated structure combining tilts and off-center displacements of Ti along the rotation axis~\cite{Goian2012, Kim2013}. This proposition however was not confirmed by structural refinements, and no such modulation was found under pressure~\cite{bes2018a}. Much more relevant to the present case are the modulations observed in the model antiferroelectric perovskites PbZrO$_3$ and PbHfO$_3$ under pressure~\cite{Burkovsky2017,Burkovsky2019} and recently fully refined in PbHfO$_3$ as the intermediate phase bridging the paraelectric cubic phase and the PbZrO$_3$-type commensurate antiferroelectric phase~\cite{Bosak2020}. The modulation in PbHfO$_3$ shares some similarities with KNbO$_3$ in that it involves displacements of the $A$ cation as well as additional octahedra tilts. On the other hand, its structure was refined with a higher orthorhombic $Imma(00\gamma)s00$ symmetry, its modulation vector was found to be only weakly temperature-dependent, the intermediate $I4/mcm$ commensurate tilted phase is also not present in PbHfO$_3$ and the transition mechanism was discussed as triggered rather than soft-mode driven~\cite{Burkovsky2019}, all of which suggest significant differences in the transition mechanisms.  

Generally, incommensurate modulations are understood as the result of a tight competition between different instabilities~\cite{Cummins1990}. Here it makes no doubt that the competition between octahedra tilts and a polar instability is the origin of the observed modulation, which in turns demonstrate the re-entrance of the polar instability. The idea of this competition is not new; its basic understanding was substantiated very early on from first principles~\cite{Zhong1995}. However, experimentally, it had led so far to situations where both instabilities are either mutually exclusive, as exemplified by PbTiO$_3$ where the polar instability needs to be killed first by hydrostatic pressure before the tilt instability can develop~\cite{Janolin2008}, or cause transitions to complex structures by convoluted transition mechanisms like in PbZrO$_3$~\cite{Tagantsev2013}. The detailed conditions that allow this original phase transition sequence in KNbO$_3$ will require further theoretical and experimental investigations, but a couple of basic observations can be made. First of all, KNbO$_3$ has a much larger tolerance factor than PbTiO$_3$ (1.06 vs. 1.03) which gives the polar instability a head start under pressure. Besides, it is remarkable that this situation is found in a compound with a significant order-disorder character, while other attempts with more displacive compounds were unsuccessfull. This raises the question whether order-disorder might in fact favor such a coexistence. We also note that the modulation in KNbO$_3$ demonstrates that it does not depend on the presence of Pb or a cation with a lone electron pair. Finally, an obvious open question is whether or not another transition happens in KNbO$_3$ at even higher pressures above \SI{63}{\GPa}. Even though the modulation vector tends towards $\nicefrac{3}{4}$, we anticipate that a true lock-in transition towards a 4-fold superstructure cannot be realized without breaking the $I$-centering since 4\textbf{q} corresponds to a $(030)$ reflection that is forbidden. Instead, it is tempting to hypothesize that KNbO$_3$ will undergo a transition to a centrosymmetric, antiferroelectric phase similar to PbHfO$_3$ and PbZrO$_3$. This will have to be confirmed experimentally. First-principle calculation will also be instrumental in checking the stability of those phases under pressure.


In summary, we have shown that KNbO$_3$ exhibits evidence for the re-emergence of a polar instability in the high-pressure regime. Due to the competition with the tilt instability, this polar instability does not give rise to a ferroelectric phase but to a modulated centrosymmetric structure that combines octahedra tilts and polar displacements of both $A$ and $B$ cations. Whether or not a macroscopically polar phase can appear at even higher pressures, in KNbO$_3$ or in other systems, remains an open question. This calls for an in-depth re-investigation of the pressure-temperature phase diagram of KNbO$_3$, especially towards the low-temperature region where we might expect an even more intricate competition of ferroelectricity and tilts. The conditions that make this possible in KNbO$_3$, and specifically the possible role of some order-disorder in balancing the competition between instabilities, will also require theoretical investigations with and beyond standard DFT approaches.

{\it Acknowledgments -- }The authors thank Philippe Ghosez and Hao-Cheng Thong for helpful discussions on preliminary first-principle calculations as well as Jeroen Jacobs for the preparation of the high-pressure cells and Stany Bauchau for technical support at the ESRF beamline. We acknowledge SOLEIL and the ESRF for provision of synchrotron radiation facilities under proposals N.~20231689, 20241276 and HC-6148 respectively.  S.R. and P.R. thank the support from the Agence Nationale de la Recherche under the project SUPERNICKEL (Grant No.ANR-21-CE30-0041-04). We also acknowledge the Interdisciplinary Thematic Institute 2021-2028 and EUR QMat (ANR-17-EURE-0024), as part of the ITI 2021-2028 program supported by the IdEx Unistra (ANR-10-IDEX-0002) and SFRI STRAT’US (ANR-20-SFRI-0012) through the French Programme d’Investissement d’Avenir. For the purpose of open access, and in fulfillment of the obligations arising from the grant agreement, the author has applied a Creative  Commons Attribution  4.0  International  (CC BY 4.0) license  to  any  Author Accepted Manuscript version arising from this submission.

The high-pressure X-ray diffraction data is available at https://doi.esrf.fr/10.15151/ESRF-ES-2107750542.\\

\bibliography{main}

\end{document}


\newcommand{\UniLu}{University of Luxembourg, Department of physics and materials science, 41 rue du Brill, 4422 Belvaux, Luxembourg}
\newcommand{\LIST}{Materials Research and Technology Department, Luxembourg Institute of Science and Technology (LIST), 41 rue du Brill, 4422 Belvaux, Luxembourg}
\newcommand{\ULiege}{Quantum Materials Center (Q-MAT), Complex and Entangled Systems from Atoms to Materials (CESAM), Universit\'e de Li\`ege, Quartier Agora, All\'ee du Six Aout, B-4000 Li\`ege, Belgique}
\newcommand{\Neel}{Institut N\'eel CNRS/UGA UPR2940, 25 Rue des Martyrs, 38042 Grenoble, France}
\newcommand{\Soleil}{Synchrotron SOLEIL, L’Orme des Merisiers, Départementale 128, 91190 Saint-Aubin, France}
\newcommand{\CNR}{CNR-SPIN, c/o Universit\`a ``G. d'Annunzio", Chieti, Italy}
\newcommand{\ILM}{Institut Lumi\`ere Mati\`ere (ILM)-UMR 5306, Universit\'e Claude Bernard Lyon 1, Campus LyonTech-La Doua, 10 rue Ada Byron, F-69622 Villeurbanne, France}
\newcommand{\ICMCB}{Institut de Chimie de la Mati\`ere Condens\'ee de Bordeaux (ICMCB)-UMR 5026, CNRS, Universit\'e de Bordeaux, 87 Avenue du Docteur Schweitzer, F-33608 Pessac, France}
\newcommand{\ESRF}{ESRF -- European Synchrotron Radiation Facility, 38000 Cedex Grenoble, France}
\newcommand{\ALBA}{CELLS -- ALBA Synchrotron Light Source, Cerdanyola del Valles, Barcelona E-08290, Spain}
\newcommand{\ICPMS}{Institut de Physique et Chimie des Matériaux de Strasbourg, UMR 7504, Université de Strasbourg, CNRS, 23 rue du Loess 67034 Strasbourg, France}


\title{An underdog story: Re-emergence of a polar instability at high pressure in KNbO$_3$\\[24pt]Supporting Information}

\author{Mohamad Baker Shoker}
\thanks{These authors contributed equally to this work.}
\affiliation{\UniLu}
\author{Sitaram Ramakrishnan}
\thanks{These authors contributed equally to this work.}
\affiliation{\Neel}
\author{Boris Croes}
\author{Olivier Cregut}
\author{Nicolas Beyer}
\author{Kokou D. Dorkenoo}
\affiliation{\ICPMS}
\author{Pierre Rodi\`ere}
\affiliation{\Neel}
\author{Bj\"orn Wehinger}
\author{Gaston Garbarino}
\author{Mohamed Mezouar}
\affiliation{\ESRF}
\author{Marine Verseils}
\author{Pierre Fertey}
\affiliation{\Soleil}
\author{Salia Cherifi-Hertel}
\affiliation{\ICPMS}
\author{Pierre Bouvier}
\affiliation{\Neel}
\author{Mael Guennou}
\affiliation{\UniLu}

\date{\today}


\maketitle


\clearpage

\section{High-pressure setup}

KNbO$_3$ single crystals grown by the top seeded solution growth were purchased from SurfaceNet GmbH (Germany) and polished down to \SI{10}{\um} for high-pressure experiments. Three different crystals of size 20x20 \si{\um} have been extracted from the parent crystal wafer. 

High-pressure experiments were performed using  membrane-driven diamond-anvil cells (DACs) with 250/300 \si{\um} bevelled diamond culets.  Rhenium gasket were used for the pressure chambers. The pressure transmitting medium used was Helium, loaded at 1.2 kbar, to ensure high hydrostatic pressure conditions up to the highest pressure reached in this study. The pressure was measured using the R1-line emission of a ruby ball placed close to the sample using the International Practical Pressure Scale IPPS-Ruby2020 equation of state~\cite{Shen2020}. The ruby signal is measured before and after each measurement in order to control the pressure drift during long acquisitions. The recorded pressure is set at the average of these two pressure values and the uncertainty is set as the half of the difference between these two values. It is typically found smaller than the symbol size used for the figures in this paper. The homogeneity of the pressure in the DAC was followed from both the width and the splitting between the R1 and R2 ruby lines~\cite{Takemura2001,Dewaele2007}.

\section{X-ray diffraction}

\subsection{Experimental details at ID27 and ID15B}

Single crystal X-ray diffraction (XRD) experiment was done at ID15B beamline (ESRF Grenoble) with a monochromatic wavelength $\lambda =\SI{0.41020}{\angstrom}$ and a $2\times 4\si{um}$ focused beam. Diffraction images were collected during the continuous rotation of the DAC around the vertical $\omega$ axis in a range ±32°, with an angular step of $\Delta\omega =\SI{0.5}{\degree}$ and an exposure time of 0.5 s/frame. The CrysAlisPro software package \cite{crysalis} was used for the analysis of the single-crystal XRD data (indexing, data integration, frame scaling, and absorption correction). A single crystal of Vanadinite [Pb$_5$(VO$_4$)$_3$Cl, $Pbca$ space group, $a = 8.8117(2)$ \AA{}, $b = 5.18320(10)$ \AA{}, and $c = 18.2391(3)$ \AA{}] was used to calibrate the instrumental model in the CrysAlisPro software, i.e., sample-detector distance, detector’s origin, goniometer angles offsets, rotation of both the X-ray beam and detector around the instrument axis. 

\subsection{Superspace approach to tackle the incommensurate phase}

From initial observation of the diffraction data as shown in a histogram from CrysAlisPro \cite{crysalis}, it was thought the system could be (3+2)-$d$ in the modulated phase as there were satellites along both \textbf{a*} and \textbf{b*} at $\approx$ 0.8. However, the (3+1)-$d$ setting was favored over the (3+2)-$d$ as there were no mixed order satellites involving a linear combination of the wavevectors as they are from different domains
related by the loss of the 4-fold rotational symmetry.

\begin{figure}[ht]
    \centering
   \includegraphics[width=80mm]{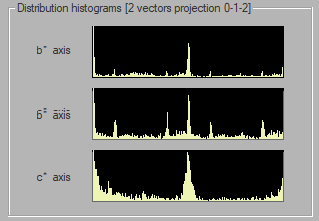}
    \caption{Histogram that depicts main and satellite peaks from CrysAlisPro. Each panel corresponds to different axis in reciprocal space. The sharp intense peaks are the main Bragg reflections, while the surrounding less intense peaks are the superlattice reflections at incommensurate positions.}
    \label{fig:ptlattice}
\end{figure}

Ergo, integration of the intensities followed by subsequent averaging and 
scaling of the various pressure points
was performed with the software CrysAlisPro \cite{crysalis}.
For each data set,
this resulted in lattice parameters,
components of the modulation wave vector
(for the incommensurate phase at 58 GPa) and
a list of integrated intensities of Bragg reflections.
Above the pressure point 44 GPa in the incommensurate phase, 
no significant deviations from tetragonal
symmetry ($I$4/$mcm$) could be detected.

Structure refinements have been performed
with the software {Jana2020} \cite{petri2023a} against the integrated intensities of Bragg
reflections .
Using the superspace approach as described in \cite{van2007incommensurate} 
the intensities of main Bragg reflections
were distinguished from the superlattice reflections
by treating it as (3+1)D, where
the displacement modulation is described by
a modulation function for each atom of the form
\begin{equation}
\mathbf{u}(\bar{x}_{s4}) =
(u_{x}(\bar{x}_{s4}),\,u_{y}(\bar{x}_{s4}),\,
u_{z}(\bar{x}_{s4})) \, ,
\label{eqsm:sral4_modulation_function_def}
\end{equation}
where for $\alpha = x, y, z$,
\begin{equation}
u_{\alpha}(\bar{x}_{s4}) = \sum_{n=1}^{n_{max}}\,\{
A_{n,\alpha} \sin(2\pi n\bar{x}_{s4}) +
B_{n,\alpha} \cos(2\pi n\bar{x}_{s4}) \} \, .
\label{eqsm:sral4_modulation_function_harmonics}
\end{equation}
We have used $n_{max} = 1$ (up to first-order harmonics).
Table \ref{tab:KNO3_I2c_ampl} shows the displacement amplitudes for every atom.

\begin{table}[ht]
\begin{center}
\small
\caption{\label{tab:KNO3_I2c_ampl}%
Amplitudes of the modulation functions of crystal KNO1 at 58 GPa, 300 K for superspace group $I$2/$c$(0\,$\sigma_1$\,$\sigma_2$)0$s$.
Values of modulation amplitudes have been multiplied
by the corresponding lattice parameter,
in order to obtain values in \AA{}.}
\begin{ruledtabular}
\begin{tabular}{ccccc}
Atom                  & K & Ni &  O1 & O2 \\
$A_{1,x}\, a$ (\AA{}) & 0.0297(13) & 0.0276(5) & -0.0002(45) & 0.173(4) \\
$A_{1,y}\, b$ (\AA{}) & 0          & 0.1572(13)           & 0           & 0.148(3) \\
$A_{1,z}\, c$ (\AA{}) & 0          & 0.004(16)           & 0           & 0.07(5) \\
$B_{1,x}\, a$ (\AA{}) & 0          & 0           & 0           & 0.003(3) \\
$B_{1,y}\, b$ (\AA{}) &0.04(3) &0 &-0.02(3) &0.101(4) \\
$B_{1,z}\, c$ (\AA{}) & -0.166(8)          & 0           & 0.10(2)           & -0.14(5) \\
\end{tabular}
\end{ruledtabular}
\end{center}
\end{table}

\clearpage

\subsection{Validation of monoclinic symmetry in the incommensurate phase}

On basis of statistics primarily the fit to the data and the data-to-parameter ratio, we have chosen to describe the modulation with the monoclinic superspace group $I$2/$c$(0\,$\sigma_1$\,$\sigma_2$)0$s$ which is essentially a subgroup
of $Ibam(0\,\sigma\,0)s00$. Table \ref{tab:KNbO3 comparemod} describes the comparison in detail.

\begin{table}[ht]
\centering
\caption{\label{tab:KNbO3 comparemod}%
Comparison of the orthorhombic and monoclinic refinements for the incommensurate phase at $P = \SI{58}{\GPa}$. Criterion of observability: $I>3\sigma(I)$. Only domain 1 reflections and common reflections have been used for the comparison. Anisotropic atomic displacement parameters are not constrained by the tetragonal $I$4$/mcm$ symmetry or disabled.}
\centering
\begin{ruledtabular}
\begin{tabular}{ccc}
Superspace group  &$Ibam(0\,\sigma\,0)s00$ & $I$2/$c$(0\,$\sigma_1$\,$\sigma_2$)0$s$  \\
$R_{int}$ $(m = 0)$ (obs/all) & 0.0158/0.0160  &0.0054/0.0055  \\
$R_{int}$ $(m = 1)$ (obs/all)   &0.0513/0.0560  & 0.0452/0.0469  \\
$R_{F}$ ($m = 0$) (obs/all)  & 0.0157/0.0174    & 0.0189/0.0226    \\
$R_{F}$ ($m = 1$) (obs/all)  & 0.0603/0.0809    & 0.0460/0.0624   \\
Parameters & 28   & 44     \\
Unique ($m = 0$) (obs/all)   & 65/84  & 135/177   \\
Unique ($m = 1$) (obs/all)   & 72/165   & 189/259  \\
\end{tabular}
\end{ruledtabular}
\end{table}

\clearpage

\subsection{Crystallographic table of various phases}

Tables \ref{tab:KNbO3_crystalinfo} shows the crystallographic information at various pressure points at room temperature indicating the different phases for the different crystals. Successful refinements were performed for all periodic phases as well as the incommensurate phase at 58 GPa. Atomic coordinates for the two new phases ($I$4/$mcm$ and $I$2/$c$(0\,$\sigma_1$\,$\sigma_2$)0$s$) are given in Table \ref{stab:KNbO3_atomic}. Data was collected at two different high pressure beamlines Id27 and Id15B in ESRF Grenoble, France.

\begin{table}[ht]
\caption{\label{tab:KNbO3_crystalinfo}%
Crystallographic data of crystal
KNbO$_3$ at \SI{300}{\K} at various pressure points for each phase. The criterion of observability is $I>3\sigma(I)$. The incommensurate
phase was collected at ID15B, while the periodic phases were collected at the ID27 beamline.}
\scriptsize
\centering
\begin{ruledtabular}
\begin{tabular}{cccccc}
Pressure (GPa) & 0  & 9 & 13 & 38.8 & 58  \\
Crystal name  & KNO3    & KNO2   & KNO2   & KNO3 & KNO1 \\
Crystal system & Orthorhombic & Tetragonal & Cubic & Tetragonal & Monoclinic \\
Space/Superspace group & $Amm$2  &$P$4$mm$ & $Pm\bar{3}m$ & $I$4/$mcm$ & $I$2/$c$(0\,$\sigma_1$\,$\sigma_2$)0$s$   \\
Space/Superspace group No. \cite{stokesht2011a}  & 38 & 99 & 221 & 140 & 15.1.4.1 \\
$a$ (\AA{}) &4.0396(6) & 3.932(3) & 3.912(2) & 5.3680(3) & 5.1972(3)              \\
$b$ (\AA{}) &5.663(4)    & 3.932 & 3.912  & 5.3680 & 5.2053(4)                \\
$c$ (\AA{}) &5.689(4)  & 3.926(2) & 3.912 & 7.564(6) & 7.552(3)                \\
$\alpha$ (deg) &90  & 90 & 90 & 90 & 89.40(10)               \\
Volume (\AA{}$^3$) & 130.2(2) & 60.7(1) & 59.8(1) & 218.0(2) & 204.3(2)         \\
Wavevector \textbf{q} & --&-- &-- &-- &0.7717(19)\textbf{b*} \\
$Z$ & 2 & 1 & 1 & 4 & 4 \\
Wavelength (\AA{}) & 0.3738 & 0.3738 & 0.3738 & 0.3738 & 0.4099    \\
Detector distance (mm) &185 & 185 &185 &185 &170.85   \\
$\omega$-scans (deg) &-32 to +32  &-32 to +32  &-32 to +32 &-32 to +32 &-32 to +32   \\
Rotation per image (deg) & 0.5 & 0.5 & 0.5 & 0.5 & 0.5   \\
$(\sin(\theta)/\lambda)_{max}$ (\AA{}$^{-1}$) &0.905659& 0.899200 & 0.930604 & 0.931447 & 0.881323 \\
Absorption, $\mu$ (mm$^{-1}$) & 5.191 & 4.9250 & 5.644 & 6.199 & 8.464   \\
No. of reflections measured, \\
$(m = 0)$  &  149  & 114 & 43 & 240 & 550\\
$(m = 1)$  & --     & -- & -- & -- & 1035 \\
No. of unique reflections,   \\
$(m = 0)$ (obs/all) & 77/77 & 60/61 & 21/22 & 41/58 & 184/238 \\
$(m = 1)$ (obs/all) & -- & -- & -- & -- & 399/518 \\
$R_{int}$ $(m = 0)$ (obs/all) &0.0599/0.0599 & 0.0826/0.0826 & 0.0544/0.0545 & 0.0260/0.0261 & 0.0443/0.0444   \\
$R_{int}$ $(m = 1)$ (obs/all) &-- & -- & -- & -- & 0.0451/0.0466 \\
No. of parameters &13 & 9 & 5 & 8 & 36 \\
$R_{F }$ $(m = 0)$  (obs) &0.0638 & 0.0492 & 0.0551 & 0.0357 & 0.0282\\
$R_{F }$ $(m = 1)$ (obs) &-- & -- & -- & -- & 0.0663\\
$wR_{F }$ (all)  &0.0751 & 0.0684 & 0.0791 & 0.0537 & 0.0606  \\
GoF (obs/all) &4.05/4.05 & 6.25/6.19 & 8.08/7.83 & 4.55/3.80 & 2.50/2.33\\
$\Delta\rho_{min}$, $\Delta\rho_{max}$(e \AA$^{-3}$) &
 2.45, -2 & 2.59, -1.91 & 2.51, 1.85 & 0.7, -0.67 & 2.72, -2.47 \\
\end{tabular}
\end{ruledtabular}
\end{table}


\begin{table}[ht]
\scriptsize
\caption{\label{stab:KNbO3_atomic}%
Atomic coordinates $x$, $y$, $z$ and
atomic displacement parameters (ADPs) $U_{ij}$
for the $I$4/$mcm$ and the $I$2/$c$(0\,$\sigma_1$\,$\sigma_2$)0$s$ incommensurate phase of KNbO$_3$. At higher pressure points above \SI{30}{\GPa}, the anisotropic ADP's of oxygen could not be refined successfully, hence they are constrained to isostropic.}
\begin{ruledtabular}
\begin{tabular}{lcccccccccc}
Atom & $x$ & $y$ & $z$ & $U_{11}$ & $U_{22}$ & $U_{33}$ & $U_{12}$ & $U_{13}$ & $U_{23}$ & $U^{eq}_{iso}$ \\
\hline
\multicolumn{11}{l}{Basic structure of KNbO$_3$ at P $= 37$ GPa, SG: $I$4/$mcm$}  \\
\hline
K  & 0 & 0.5 & 0.25 & 0.017(1) &  0.017 & 0.03(3)& 0 & 0 & 0 & 0.024(10) \\
Nb &0  & 0 &  0 & 0.0147(5)&  0.0147(3)& 0.107(2)& 0 & 0 & 0 & 0.045(7) \\
O1 &0  & 0 &  0.25 & - & - & - & - & - & - & 0.09(4) \\
O2  & 0.2417(8) & 0.7417 & 0 & - & - & - & - & - & - & 0.019(2) \\
\hline
\multicolumn{11}{l}{Basic structure of KNbO$_3$ at P $= 58$ GPa, SSG: $I$2/$c$(0\,$\sigma_1$\,$\sigma_2$)0$s$} \\
\hline
K  & -0.0005(8) & 0.5 & 0.25 & 0.0101(10) &  0.009(2) & 0.027(10)& 0 & 0 & 0.001(8) & 0.0015(4) \\
Nb &0  & 0 &  0 & 0.0082(3)&  0.0045(3)& 0.030(4)& -0.0009(2) & -0.0002(12) & -0.002(5) & 0.014(1) \\
O1 &0.008(3)  & 0 &  0.25 & - & - & - & - & - & - & 0.014(2) \\
O2  & 0.2058(5) & 0.7038(6) & 0.017(4) & - & - & - & - & - & - & 0.006(1) \\
\end{tabular}
\end{ruledtabular}
\end{table}

\clearpage

\subsection{Birch-Murnaghan fit to the cubic volume}

\begin{figure}[ht]
    \centering
   \includegraphics[width=100mm]{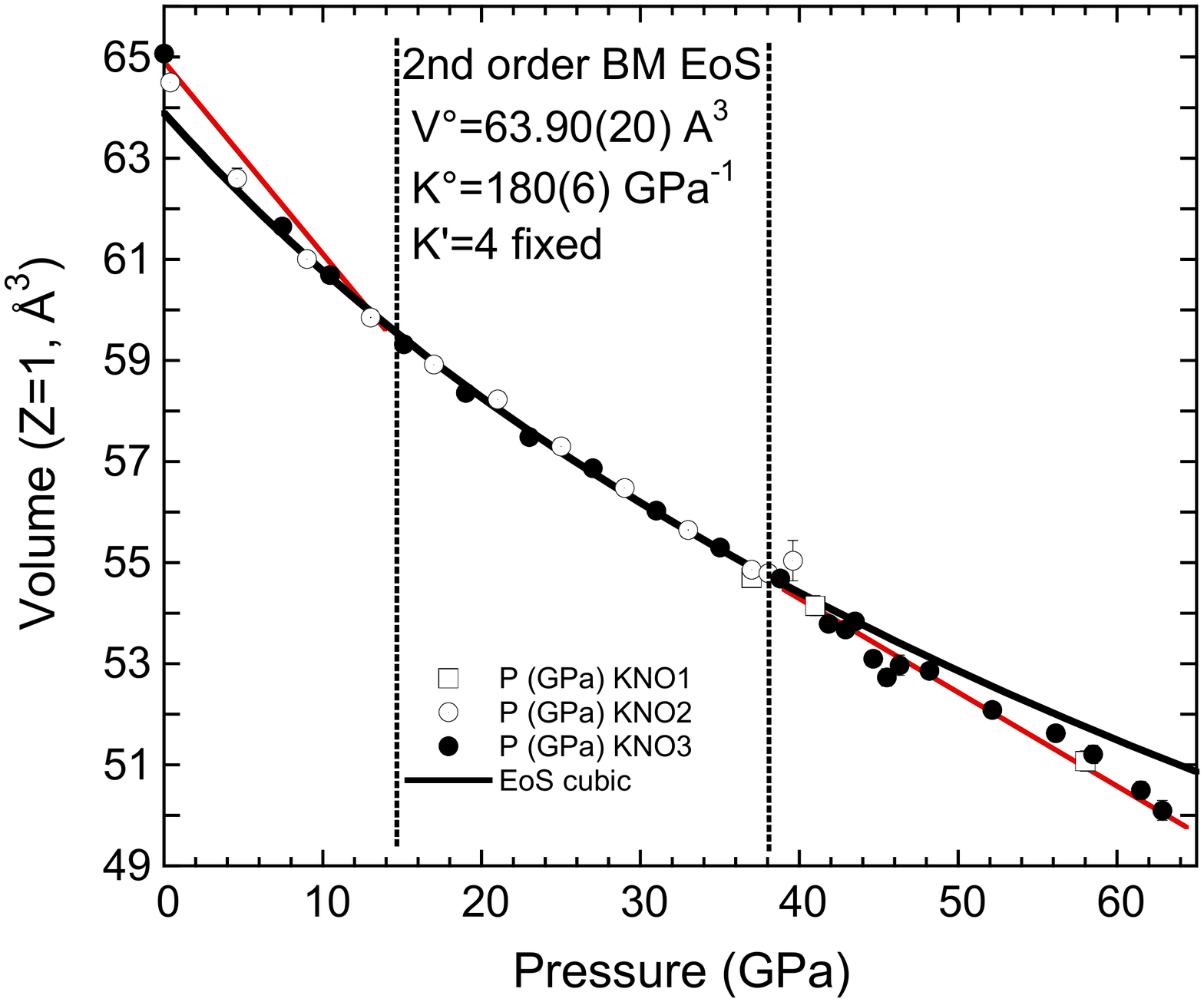}
    \caption{Pressure evolution of the pseudo-cubic volume with 2\textsuperscript{nd}-order Birch-Murnaghan equation of state (BM EoS).}
    \label{fig:EoS}
\end{figure}

The pressure dependence of the cubic volume is fitted with a second-order Birch-Murnaghan equation of state~\cite{Birch1947} (BM EoS) using equation \ref{eqEos} with volumes between 15 and \SI{38}{\GPa} only. At zero pressure, the BM EoS parameters are $V^{\circ}=63.90(20)$ \AA{}$^3$, $K^{\circ}=\SI{180(6)}{\GPa}$ and $K^{\prime}=4$ (fixed). Figure \ref{fig:EoS} shows this equation extrapolated over the entire pressure range explored, in the ferroelectric phases ($P\leq \SI{15}{\GPa}$) and in the high-pressure phases ($P \geq\SI{38}{\GPa}$). It can be seen that the low-pressure ferroelectric pseudo-cubic volume is slightly larger, whereas the high-pressure pseudo-cubic volume is slightly smaller than the cubic one. This validates the general principle that, at high-pressure, the octahedral tilt instability reduces the overall volume of the perovskite.

\begin{equation}
P=\frac{3}{2}K^{\circ}[(V_{\circ}/V)^{7/3}-(V_{\circ}/V)^{5/3}]
\label{eqEos}
\end{equation}

\clearpage



\clearpage
\section{Second Harmonic Generation}

Second Harmonic Generation (SHG) measurements were performed using an inverted optical microscope equipped with a mode-locked laser that delivered 100-\si{\fs} pulses at an 80-\si{\MHz} repetition rate. The sample was illuminated at normal incidence with a \SI{800}{\nm} wavelength and a time-averaged power of approximately \SI{20}{\mW}. To ensure power stability throughout the measurements, the incident laser power was recorded before and after each acquisition using a calibrated power meter. Two-dimensional image acquisition was performed by scanning the sample relative to the laser focus using computer-controlled stepping motors. Optical focusing was achieved using a long-working-distance $40\times$ objective with a numerical aperture of $0.4$. The SHG signal was spectrally filtered and detected using a photomultiplier tube. Further technical details concerning the SHG polarimetry microscope and data analysis procedures are available in Refs.~\cite{Cherifi-Hertel2021, Croes2022}.

Ruby fluorescence was excited with a \SI{457}{\nm} laser source within the same SHG microscope using an retractable mirror to alternate the optical path between SHG and fluorescence detection modes. The emitted fluorescence was measured with a \SI{300}{\mm} spectrometer (SP300i, Acton Research) with a $1200$~grooves/mm grating and a \SI{100}{\um} entrance slit size.

\clearpage
\section{Raman spectroscopy}

\subsection{Experimental details}

The Raman measurements were performed in backscattering geometry with a 50X objective (Nikon) to focus the incident laser beam and collect the scattered light from inside the DAC through the diamond anvil. The spectra are measured at room temperature using a \SI{514.4}{\nm} laser (Cobolt Fandango) and a \SI{750}{\mm} spectrometer (SP2750, Acton Research) with a 1800 grooves/mm grating (blazed at \SI{500}{\nm}), equipped with a cooled CCD camera (PyLoN, Princeton), and a \SI{50}{\um} entrance slit size that provides a resolution of \SI{1.09}{\per\cm} (\SI{0.03}{\nm}). A set of Bragg filters (BNF-Optigrate) were used in order to reject the excitation line. The spectrometer was calibrated in wavenumber using the lines of a Ne-Ar lamp. The incident laser power was fixed at 8 mW (measured before the DAC). The Raman spectra covering a 15–1125 cm$^{-1}$ spectral range were recorded using two monochromator positions with a maximum of \SI{90}{\s} acquisition time averaged over three acquisitions. 

\subsection{Raman signature in the cubic phase between 14 and 38 GPa}

\begin{figure}[ht]
    \centering
    \includegraphics[width=1\textwidth]{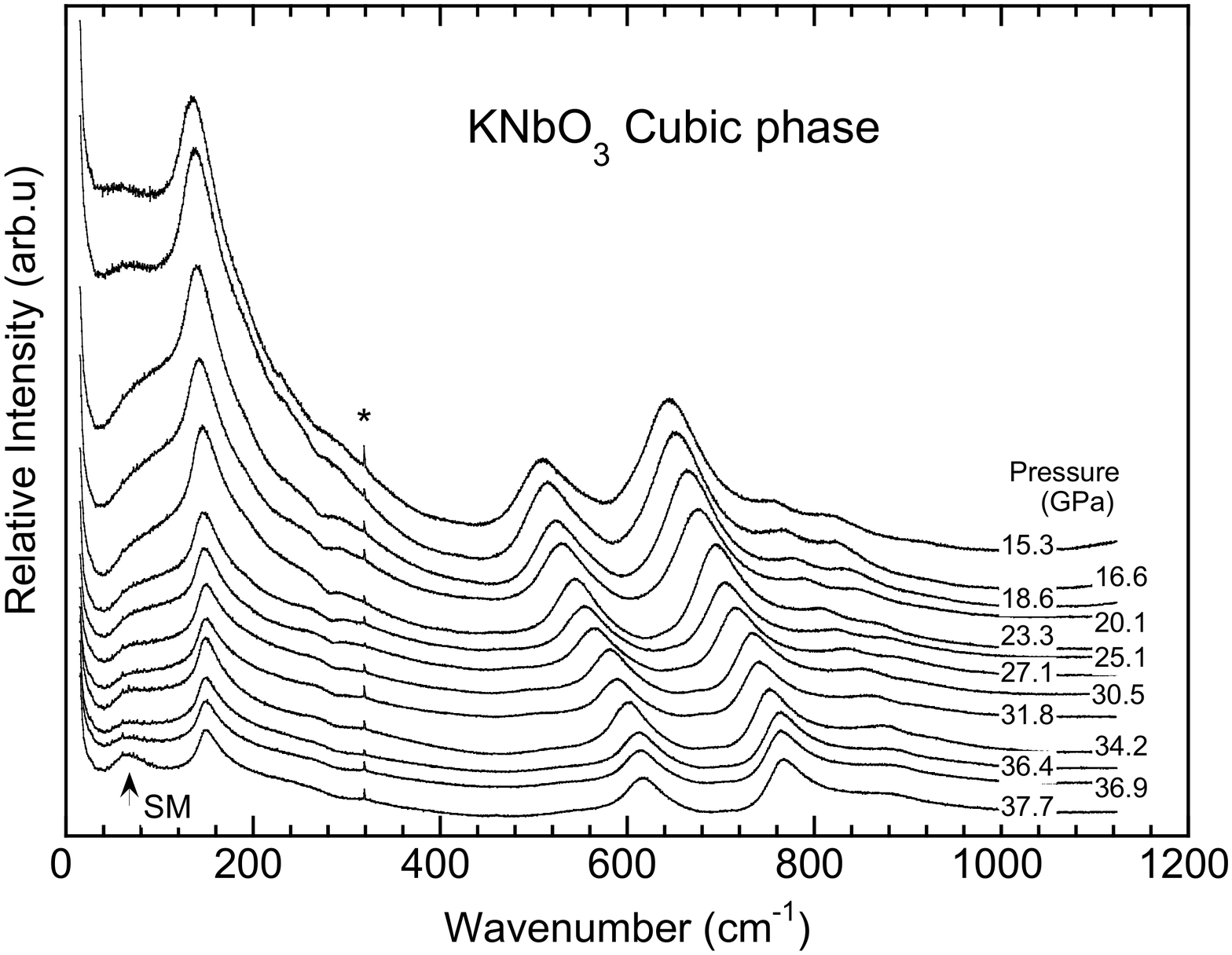}
    \caption{Raman spectra recorded in the cubic phase between 15.3 and \SI{37.7}{\GPa}. The pressure are reported in the right side of each spectrum. The star * corresponds to an artifact line from the grating. The tilt soft mode at \SI{65}{\per\cm} is labelled as SM at \SI{37.7}{\GPa}.}
    \label{fig:supp:RamanCubic}
\end{figure}

Figure~\ref{fig:supp:RamanCubic} shows the Raman sepctrum of KNbO$_3$ in its cubic phase. In the ideal cubic phase $Pm\bar 3m$ no Raman signature is expected. However, we measured a strong signal made of several broad and asymmetric bands that correspond the phonon density of state over the entire Brillouin zone. This attests that in the cubic phase the Nb are locally still displaced from the center of the octahedron. 

Also, we observe that at \SI{15}{\GPa} the tetragonal to cubic transition is associated with a strong increase of the  Rayleigh diffusion that progressively decreased after the transition up to \SI{20.1}{\GPa}.

\clearpage

\section{Fourier-transform infrared spectroscopy (FTIR)}
Infrared spectroscopy measurements in the THz and FIR energy ranges were performed in the Reflectivity configuration on an IFS125MR Michelson interferometer exploiting the synchrotron radiation extracted on the AILES beamline of synchrotron SOLEIL and the high-pressure/low-temperature setup of the AILES beamline~\cite{Voute2016}. This setup allows a strong focusing (down to \SI{100}{\um}) and a quasi-normal incidence of the synchrotron beam onto the sample surface. Measurement in the THz (10 to \SI{70}{\per\cm}) and the FIR (60 to \SI{700}{\per\cm}) energy range were performed at each pressure by using two external detectors, a \SI{1.6}{\K} pumped bolometer and a \SI{4.2}{\K} bolometer, respectively and using the same 6-\si{um} Mylar multilayer beamsplitter. The references used for the calculation of the reflectivity spectra is a piece of gold foil of the same size than the sample and placed in the hole of the gasket after the measurement of the sample. 

\clearpage
\section{First-principle calculations}

\subsection{Computational details}

Density functional theory (DFT) calculations were performed at 0 and 40~GPa using the projector augmented wave (PAW) method, as implemented in the Vienna \textit{Ab initio} Simulation Package (VASP) code. Exchange--correlation effects were treated with the PBEsol (Perdew--Burke--Ernzerhof functional revised for solids) functional under the generalized gradient approximation. A plane wave basis set with a kinetic energy cutoff of 500~eV was employed throughout.
Structural relaxations and force evaluations were performed on the primitive five-atom cubic unit cell using a $\Gamma$-centered $8 \times 8 \times 8$ \textbf{k}-point mesh to sample the Brillouin zone. Electronic self-consistency was achieved when total energy changes fell below $10^{-8}$~eV and residual forces on each atom were less than $0.001$~eV/\AA.
Phonon properties were obtained using the finite displacement approach implemented in \textsc{Phonopy}. A $2 \times 2 \times 2$ supercell of the optimized primitive cell was generated, and all symmetry-inequivalent displacements were created automatically. For each displaced configuration, single-step force calculations were carried out in separate directories. The resulting forces were collected, via \textsc{Phonopy}, to construct the interatomic force constants.

\subsection{Results}

The phonon dispersion calculated at 0 and \SI{40}{\GPa} for cubic KNbO$_3$ are given in Figure~\ref{fig:supp:DFT} in order to illustrate the hardening of the ferroelectric soft mode and the softening of the antiferrodistortive tilt mode under pressure. The modes of interest are indicated, and also described in more details in Table~\ref{tab:supp:DFT}.

\begin{figure}[ht]
    \centering
    \includegraphics[width=\textwidth]{figures_supplement/Phonon_Dispersion.pdf}
    \caption{Phonon dispersion for cubic KNbO$_3$ at 0 and \SI{40}{\GPa}.Imaginary phonon modes are plotted as negative frequencies.}
    \label{fig:supp:DFT}    
\end{figure}

\begin{table}[ht]
\centering
\footnotesize
\renewcommand{\arraystretch}{1.1}
\setlength{\tabcolsep}{10pt}
\begin{tabular}{c c l c c c}
\hline\hline
\multicolumn{2}{c}{Mode symmetry} & Displacement pattern & \multicolumn{2}{c}{DFT} & Experiment \\
in $Pm\bar 3m$ & in $I4/mcm$ &  & \SI{0}{\GPa} & \SI{40}{\GPa} &  \\
\hline
($T_{2u}$) & $(B_{2u}) + \mathbf{E_u}$ & O "butterfly" motion & 255 & 179 & Not observed \\\hline
R$_4^+$ & $A_{1g} + E_g$ & NbO$_6$ tilts  & 155 & -85 & Soft mode \\
\hline
R$_5^+$ & $B_{2g} + E_g$ & K antiparallel motion & 174 & 222 & $\sim$240 (\SI{40}{\GPa}) \\
\hline
R$_4^-$ & $(A_{1u}) + \mathbf{E_u}$ & Nb antiparallel motion  & 338 & 409 & $\sim$400 (\SI{37}{\GPa}) \\
\hline
R$_5^+$ & $B_{2g} + E_g$ & NbO$_6$ shearing & 414 & 433 & $\sim$460 (\SI{38}{\GPa}) \\
\hline
R$_3^+$  & $(A_{2g}) + B_{1g}$ & NbO$_6$ asym. stretching & 379 & 707 & Not observed \\
\hline\hline
\end{tabular}
\caption{Phonon modes of interest discussed in the text with: mode symmetries in the paraelectric and antiferrodistortive phases, atomic displacement patterns, calculated frequencies and experimental observations. The spectroscopic activity for the modes at $\Gamma$ is indicated as follows: modes written in bold are IR active, modes in brackets are silent, the others are Raman active. All phonon frequencies are in \si{\per\cm}.}
\label{tab:supp:DFT}
\end{table}

\clearpage
\bibliography{main}